\documentclass[final,english]{bullsrsl}[2022/06/15]

\usepackage[latin1]{inputenc}
\usepackage[T1]{fontenc}
\usepackage{subfigure}
\usepackage{natbib} 
\usepackage{graphicx}
\usepackage{hyperref}
\begin{document}
\title{Coronal modulation in the ultra-fast rotator LO Peg}

\author[corresponding]{Gurpreet}{Singh*$^{1,2}$}
\author[]{Jeewan Chandra}{Pandey$^1$}
     \author[]{Umesh }{Yadava$^2$}
\affiliation[]{$^1$Aryabhatta Research Institute of observational sciencES, Manora Peak, Nainital, 263001, India\\
$^2$Deen Dayal Upadhyaya Gorakhpur University, Gorakhpur, 273009, India}
\correspondance{gurpreet@aries.res.in}
\date{\date}
\maketitle

\begin{abstract}
We present coronal imaging of the ultra-fast rotator, LO Peg, using the X-ray observations from  XMM-Newton. The X-ray light curves show one strong flare at the end of observation, as reported in an earlier study. On removal of flaring events, the quiescent state light curve shows rotational modulations, which are modelled using a maximum likelihood model. The results obtained from modelling show the corona of LO Peg is not uniform. Active regions are concentrated around two longitudes, where one active region appears to be compact. The large coronal area that covers almost 60 degrees along longitude from the poles to the equator does not consist of active regions.
\end{abstract}

\keywords{ \href{http://astrothesaurus.org/uat/1580}{Stellar activity } --- \href{http://astrothesaurus.org/uat/305 }{Stellar Coronae } --- 	\href{http://astrothesaurus.org/uat/1823} {X-ray star } --- \href{http://astrothesaurus.org/uat/1810} {X-ray astronomy} --- \href{http://astrothesaurus.org/uat/1145}{Stellar imaging }}

\vspace{-0.5cm}
\section{Introduction}
The light curves of late-type active stars exhibit a wide range of temporal variability across the electromagnetic spectrum, spanning from minutes to decades. This variability can be categorized into short-term variability (STV) and long-term variability (LTV), which are manifestations of different magnetic activities.
STV, lasting from a few minutes to a few days, is primarily attributed to flaring activity and rotational modulation caused by inhomogeneities in the corona. Extensive studies and modelling of STVs due to flares have been conducted, shedding light on the extreme physical conditions of solar-type stars \citep[e.g.][]{1991ARA&A..29..275H,2007A&A...471..271R,2008MNRAS.387.1627P,2012MNRAS.419.1219P}.
Several techniques have been employed to extract information from periodic STVs resulting from rotational modulations of active regions in stellar atmospheres. These include Doppler imaging using X-ray data \citep[][]{2001ApJ...562L..75B}, extrapolation of surface magnetic maps \citep[][etc]{2007MNRAS.377.1488H,2010MNRAS.404..101J,2010ApJ...721...80C}, and light curve inversion techniques \citep[][]{1992MNRAS.259..453S,1996ApJ...473..470S,2014ApJ...783....2D,2022ApJ...934...20S}. However, each technique has its limitations.
Doppler imaging of X-ray data requires high-spectral resolution, often unavailable due to instrumental and observing constraints. Inferring coronal structures based on magnetic surface maps necessitates simultaneous observations in optical and X-ray bands. Light-curve inversion techniques (LCITs) pose a mathematically ill-posed problem, extracting 3-D information from 1-D time-series data. Despite these challenges and reasonable inputs, the LCITs have gained popularity due to the increasing availability of time-series data.

This paper presents the results obtained by an X-ray LCIT for an ultra-fast rotator (UFR) LO Peg.
LO Peg is a K5-8V type UFR with a rotation period of 0.423 d \citep[][]{2016MNRAS.459.3112K}. The active nature of LO Peg in optical and X-ray bands has been studied in detail in the past \citep[e.g. ][]{1994A&A...282L...9J,1999A&A...341..527E,2005AJ....130.1231P,2009MNRAS.396.1004P,2016MNRAS.459.3112K}.
The Doppler imaging of LO Peg has shown high-latitude spots \citep[][]{1999MNRAS.307..685L,2008MNRAS.387..237P}. Based on long-term optical data, \cite{2016MNRAS.459.3112K} has shown an excess of X-ray emission in spotted regions where the spotted regions can cover the stellar surface from  9 to 26\%. They have also found evidence of the presence of the flip-flop-like phenomenon.  

We organise our paper as follows: Section \ref{obs} shows observations and data reduction, and Section \ref{aandr} presents the analysis and results. The coronal imaging method is described in Section \ref{method}, and application to this method is shown in Section \ref{solution}. We conclude in Section \ref{conc} with the key findings from this research.

\vspace{-0.5cm}
\section{Observations and data reduction}\label{obs}
The XMM-Newton observatory observed LO Peg for 42 ks on 30 November 2014. XMM-Newton is equipped with three X-ray telescopes featuring five detectors: 2 MOS \citep{2001A&A...365L..27T}, 1 PN \citep{2001A&A...365L..18S}, and 2 RGS \citep{2001A&A...365L...7D}. These detectors cover an energy range of 0.15--15 keV. In addition to the X-ray detectors, XMM-Newton also includes an optical monitor payload \citep[OM;][]{2001A&A...365L..36M}, enabling simultaneous observations in the UV and optical bands.

The raw data obtained from the observation was processed using the Science Analysis System (SAS) v18.0.0 software (\href{https://www.cosmos.esa.int/web/xmm-newton/sas}{https://www.cosmos.esa.int/web/xmm-newton/sas}). Standard procedures provided by the software were applied to generate the science products. The source light curves were created by considering the X-ray counts within a circular region of radius 42". Similarly, background light curves were generated from source-free regions in the same CCD, utilizing an extracted area similar to the source. Subsequently, the background subtraction and detector response effects were corrected using the \textsc{epiclccorr} task.
\begin{figure}[!t]
    \centering
    \includegraphics[scale=0.4]{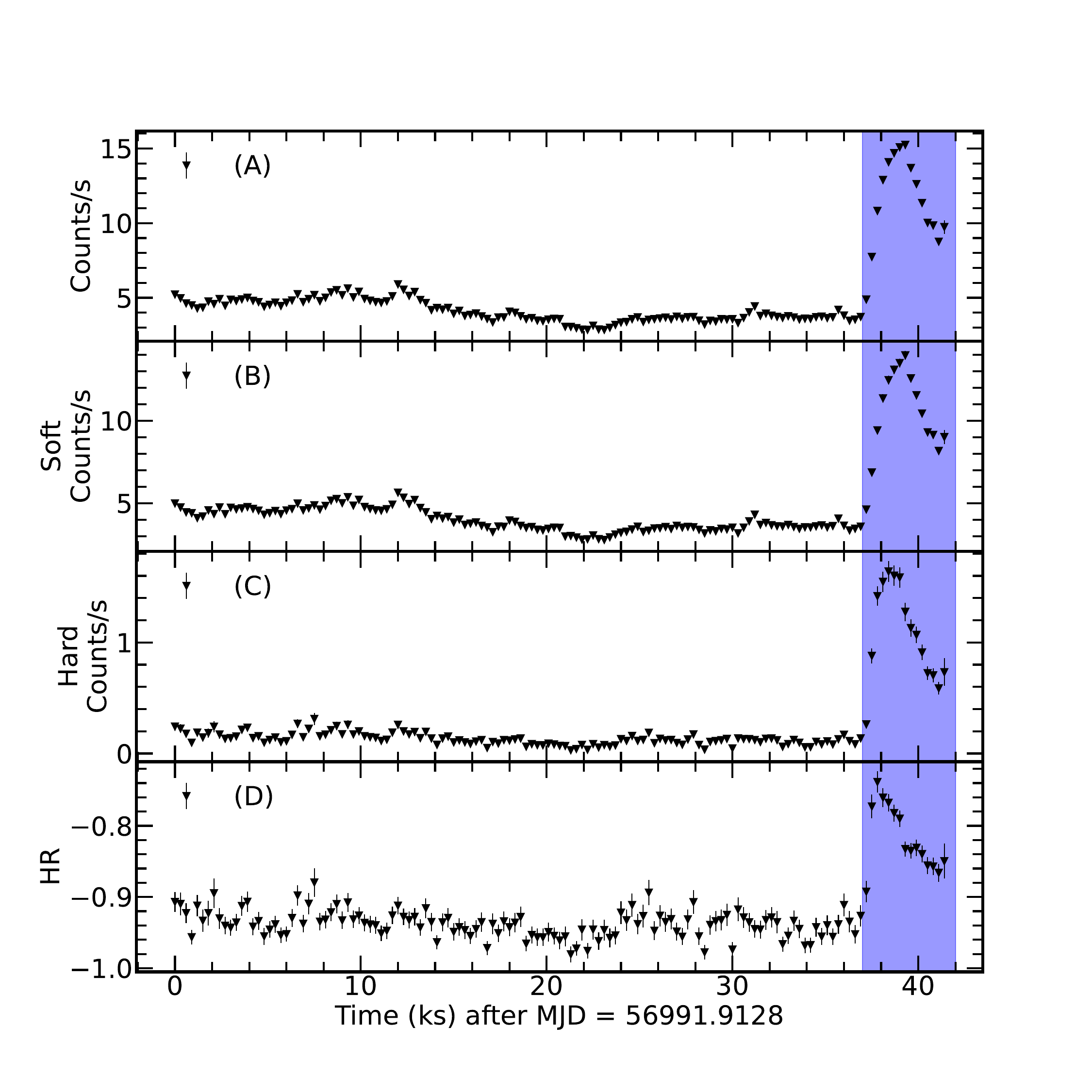}
\bigskip
\begin{minipage}{12cm}
\caption{EPIC-PN X-ray light curve of LO Peg with a time bin size of 300 sec. Panel (A) shows an X-ray light curve in 0.3--10.0 keV, with panels (B) and (C) showing soft (0.3--2.0 keV) and hard (2.0-10.0 keV) bands. The lowermost panel (D) shows the hardness ratio (HR). The blue-shaded regions correspond to flaring episodes.}
    \label{fig:1}
\end{minipage}
    
\end{figure}


\section{Analysis and results}\label{aandr}
The background corrected light curves for LO Peg are shown in Figure \ref{fig:1} in three energy bands, namely a broad band which covers 0.3-10.0 keV energy range, a soft band with an energy band of 0.3-2.0 keV, and a hard band containing photons of energy 2.0-10.0 keV.  The light curves in all three bands show one standard flare-like time profile marked as blue-shaded regions in Figure \ref{fig:1}.  This detected flare also shows temperature variations indicated by the hardness ratio (HR) plot and shown in the lowermost panel of Figure \ref{fig:1}. The HR is defined as  Hard-Soft/Hard+Soft, where  Hard and Soft are count rates in 2.0-10.0 keV and 0.3-2.0 keV energy bands, respectively.  From the HR plot, the mean value of HR $\sim$ -0.9 indicates that LO Peg's corona is predominantly populated by low-energy photons with energy less than 2 keV.  The ratio of median values of soft counts and total band counts shows that photons with energy less than 2 keV contribute 96\%  of the total band energy spectrum.  Thus, the quiescent corona of LO Peg predominantly consists of plasma with a temperature of less than 2 keV.

\begin{figure}[!t]
\centering
    \subfigure[]{\includegraphics[height=0.30\columnwidth,trim={0.0cm 1.7cm 3.0cm 3.0cm}]{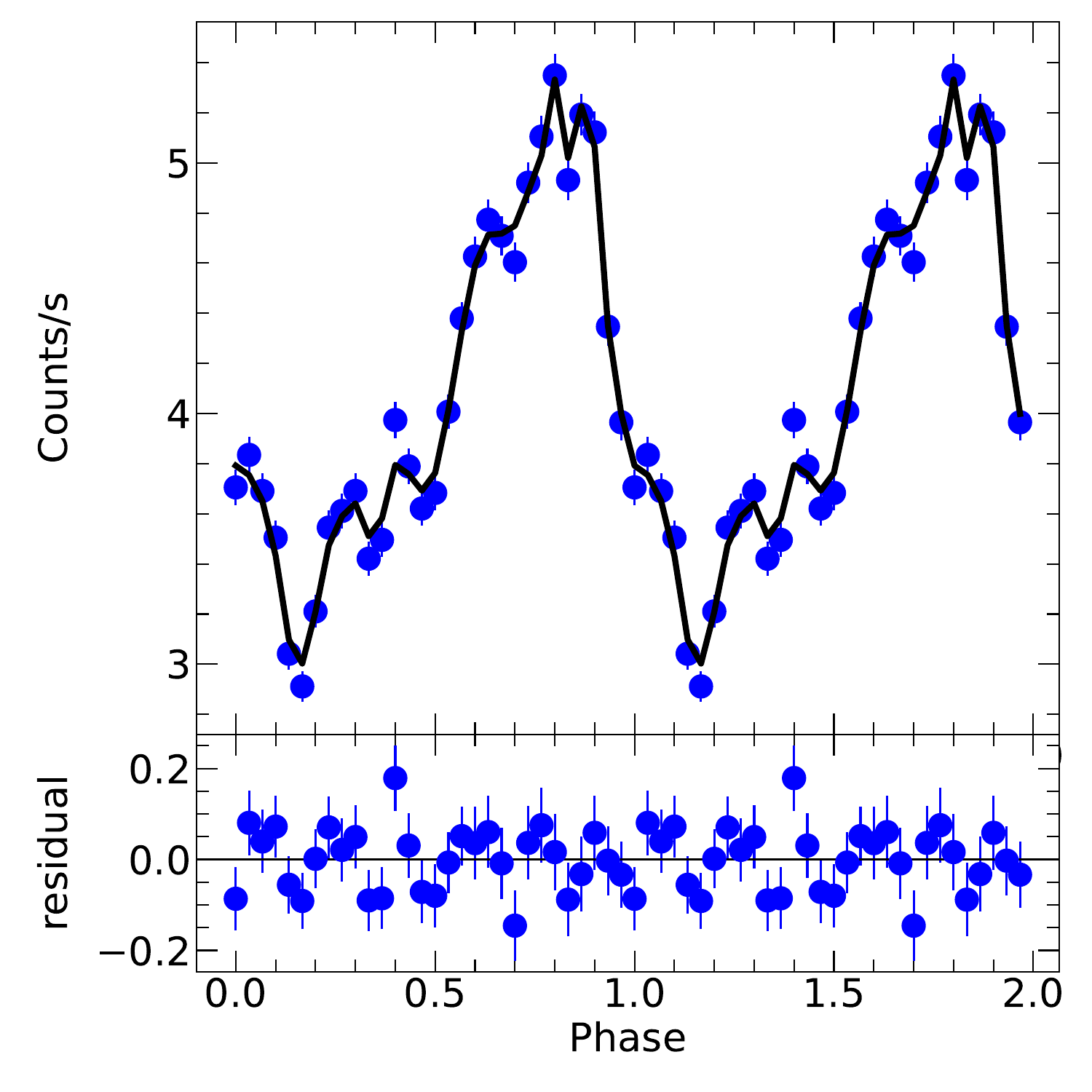} \label{fig:b}}\hspace{50pt}
    \subfigure[]{\includegraphics[height=0.36\columnwidth,trim={0.0cm 0.0cm 0.0cm 2.5cm},clip]{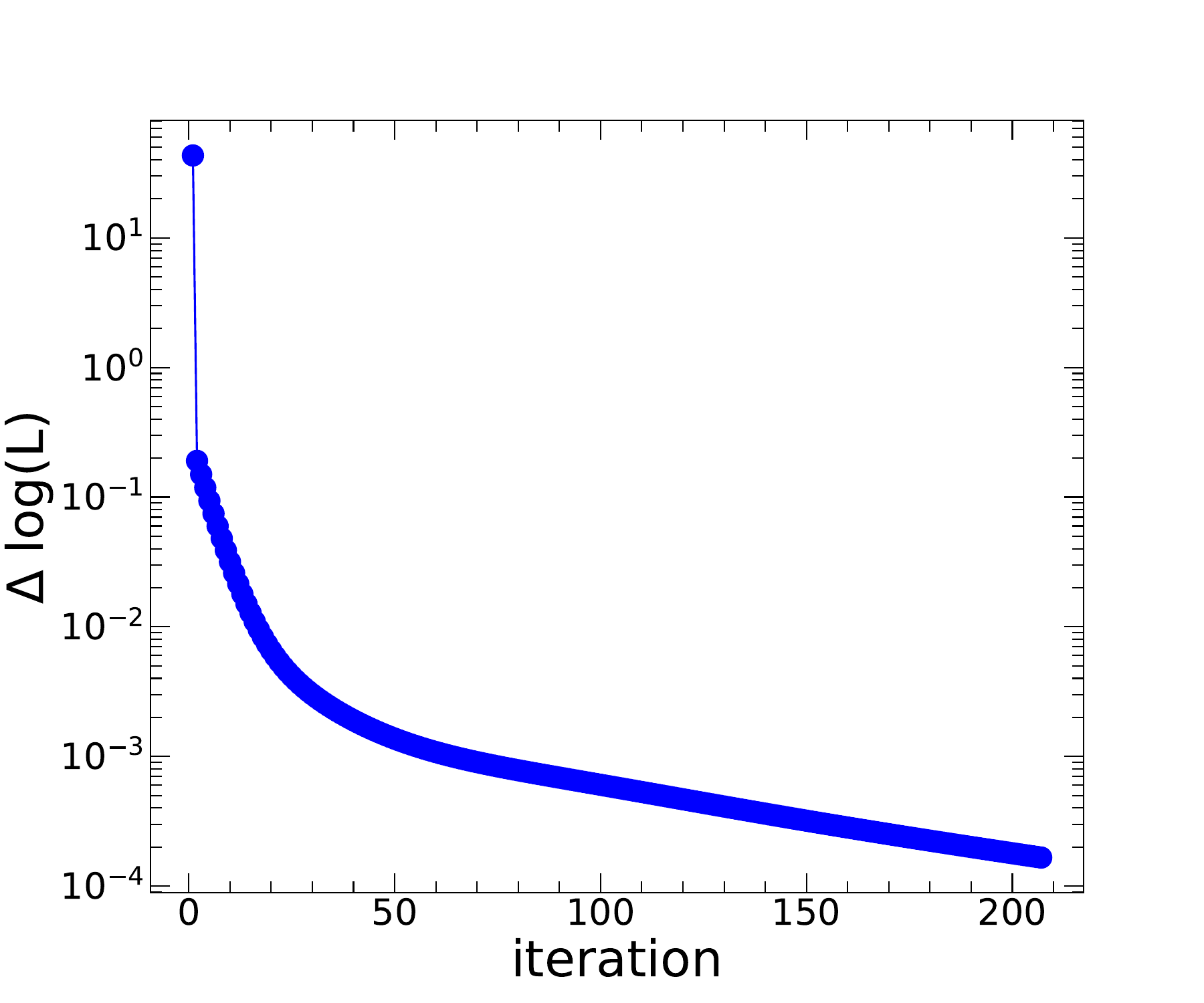} \label{fig:b}}
    \subfigure[]{\includegraphics[width=0.9\columnwidth,trim={0.0cm 1.0cm 3.0cm 0.0cm}]{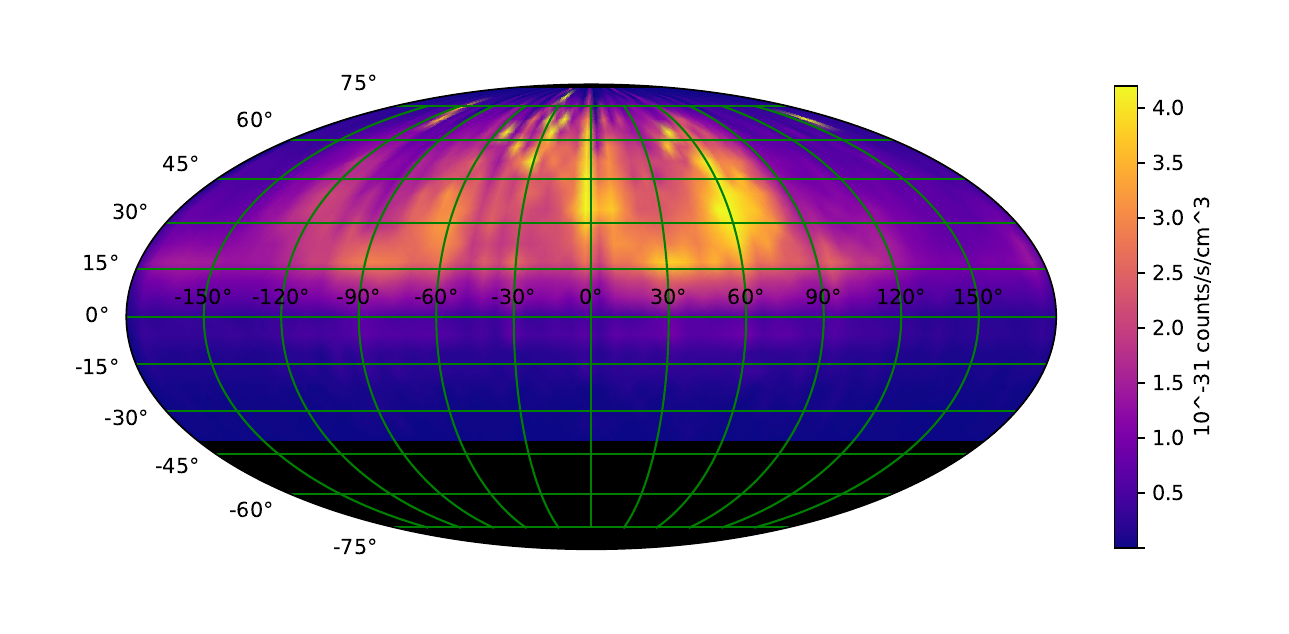}\label{fig:a}} 
        \bigskip
\begin{minipage}{12cm}
\caption{ Results from coronal imaging of LO Peg. (a) Observed and the best fit modelled X-ray light curves, (b) $\Delta log(L)$ vs iteration showing likelihood is strictly increasing in each iteration, and (c) Projected coronal image of LO Peg.  Here, the colour bar shows emission values at each latitude and longitude. The black-shaded regions correspond to latitudes which do not contribute to rotational modulation.  \label{fig:2}  }
\end{minipage}
         \end{figure}

If we exclude the flaring part from the light curve, the quiescent state light curve appears variable in the time scale of its rotation period.  Therefore, we have phase-folded the quiescent light curve to check for the rotationally modulated signal with ephemeris given by \cite{2003IBVS.5390....1D} with the rotational period obtained by  \cite{2016MNRAS.459.3112K}.
The phase folded light curve is shown in Figure \ref{fig:2} (a), and rotational modulation was found to be present with an amplitude of >20\% about the mean value.  Estimating rotational modulation from a single rotation period is prone to small-scale fluctuations.  However, it is essential to note that the rotational modulation for the same observation was previously reported by \cite{2017A&A...602A..26L}.  Additionally, to mitigate the influence of short-term stochastic variability in the folded light curve, we have phase-folded the light curve into 30 bins, corresponding to a 1.22 ks time interval.  This process allows for averaging stochastic variability caused by micro- and nano-flares, emphasising only the large-scale variations. The phase-folded quiescent light curve was modelled using a maximum likelihood algorithm, which is discussed briefly as follows. 
\section{Coronal imaging: Model}
\label{method}
We have assumed that the coronal plasma is optically thin and rotates rigidly, and the variability seen in quiescent emission is mainly due to the active regions in the corona, which are being eclipsed by the cylindrical shadow cast by the stellar photosphere during observation.  Further, we divide the volume around the star into a total of $B$ cubical boxes, each associated with an emission density $f$.  The boxes around the star are distributed up to a height $h$ above the stellar photosphere. 
The number of photons emitted by a cubical box ($b$) follows a Poisson distribution with a mean value of $f(b)$, which is given as\\
$$P(n(b)=k)=e^{-f(b)}\frac{f(b)^k}{k!} \quad ; k=0, 1, 2, ...$$
 We aim to find each box's $f(b)$. For instance, there are 'N' observations during one rotational phase, and then different parts of the star become visible with the progressing rotational phase.  As the plasma is assumed to be optically thin at any given phase value $(\phi)$, the total emission can be calculated by adding all the visible cubical boxes.  To calculate which box is visible at any given phase, we assume that the only occulter to the corona is the stellar photosphere, which casts a cylindrical shadow on the corona.  We calculate an occultation matrix $M(b,\phi)$ consisting of weights assigned to each box equal to 1 if visible and 0 if occulted.  So, the total emission at any given phase is given by 
$$F(\phi)=\sum_{b}f(b)M(b,\phi)db$$
Due to the inclination angle of the star, some boxes are never seen during the whole observation, and some boxes are always visible.  The boxes not seen should be excluded from the solution grid as they do not contribute to any of the variability seen in the light curves if we assume the emission from each box is stable during the observation.  The conditional probability of observing counts $F_o(\phi_i)$ at phase $\phi_i$ with known emissions in each box $f(b)$ and occultation matrix $M(b,\phi_i)$ can be written as
$$P_i(F_o(\phi_i)|f,M(\phi_i))=e^{-\lambda}\frac{\lambda^{F_o(\phi_i)}}{F_o(\phi_i)!};\quad \lambda=\sum_b f(b)M(b,\phi_i)db$$
The log-likelihood function can be written as
$$log(L)=\sum_i log(P_i)=\sum_i -\lambda + F_o(\phi_i)log(\lambda)- log(\Gamma(F_o(\phi_i)+1)) \quad\quad ... (1)$$

We now describe an iterative scheme that maximises the above-said setup's likelihood.
$$f^{n+1}(b)=f^n(b)\frac{\sum_i\frac{F_o(\phi_i)}{\sum_{b}f^n(b)M(b,\phi)db}M(b,\phi_i)}{\sum_iM(b,\phi_i)} \quad\quad ... (2)$$
 The above equation updates the old estimate for emission in box $b$ by considering the previous value of $f$ at point $b$, scaling it based on the ratio of $F_o(\phi_i)$ to the weighted average of $F_o$ for the all $b$ boxes, and then normalising it by the sum of the weighting function $M$ for all phases $\phi_i$.
 \section{Coronal imaging of LO Peg}
 \label{solution}
 For the current solution, we have used a resolution of 0.05$\times$0.05$\times$0.05$R_\odot$, with an inclination angle fixed to 45$^\circ$ and radius of  0.72$R_\odot$ for LO Peg \citep[][]{2005AJ....130.1231P} with coronal height upto 1 $R_*$. We start with a uniform emission in each box for the first iteration step, which is then updated according to equation (2).  At each iteration step, we also calculate the log-likelihood as per equation (1).  The iteration was stopped when observed, and modelled count rates were fitted well as per the standard $\chi^2$ scheme.  The coronal image and modelled light curves are shown in Figure \ref{fig:2}.  In the same Figure, we plot the log-likelihood function's positive difference, showing that each iteration strictly increases the likelihood. 
 
From Figure \ref{fig:2} (c), the corona of LO Peg shows active regions are not uniformly distributed across the corona. 
Active regions are located between the longitudes -120$^\circ$ to +120$^\circ$ with brighter active regions around +60$^\circ$. 
Around -90 to -60$^\circ$ longitude, the corona has a fainter active region.
Similar longitudinal distribution in the photosphere is also reported for LO Peg during the year 2014-2015 \citep[][]{2016AcA....66..381S}.  Active regions appear to be absent around the longitudes +120$^\circ$ to +180$^\circ$, which may be an indication of a coronal hole.

\section{Conclusions}\label{conc}
We have found that the quiescent emission of LO Peg is not constant but shows rotational modulation with an amplitude of >20\% about the mean value.  This modulation is modelled by a maximum-likelihood method, revealing that the corona of LO Peg is not uniform but is concentrated towards two longitudes.  Active regions are concentrated in the two different longitudes. 

\begin{acknowledgments}
This work is based on observations obtained with XMM-Newton, an ESA science mission with instruments and contributions directly funded by ESA Member States and NASA.  
\end{acknowledgments}

\begin{furtherinformation}

\begin{orcids}
\orcid{0009-0002-6580-3931}{Gurpreet}{Singh}
\orcid{0000-0002-4331-1867}{Jeewan C.}{Pandey}
\end{orcids}

\begin{authorcontributions}
All authors contributed significantly to the work presented in this paper.

\end{authorcontributions}

\begin{conflictsofinterest}
The authors declare no conflict of interest.
\end{conflictsofinterest}

\end{furtherinformation}

\bibliographystyle{bullsrsl-en}

\bibliography{main}

\begin{thebibliography}{26}
\providecommand{\natexlab}[1]{#1}
\providecommand{\url}[1]{#1}
\providecommand{\urlprefix}{URL }

\bibitem[{{Brickhouse} et~al.(2001){Brickhouse}, {Dupree} and
  {Young}}]{2001ApJ...562L..75B}
{Brickhouse}, N.~S., {Dupree}, A.~K. and {Young}, P.~R. (2001) {X-Ray Doppler
  Imaging of 44i Bootis with Chandra}.
\newblock ApJ, 562(1), L75--L78.
\newblock \url{https://doi.org/10.1086/338121}.

\bibitem[{{Cohen} et~al.(2010){Cohen}, {Drake}, {Kashyap}, {Hussain} and
  {Gombosi}}]{2010ApJ...721...80C}
{Cohen}, O., {Drake}, J.~J., {Kashyap}, V.~L., {Hussain}, G.~A.~J. and
  {Gombosi}, T.~I. (2010) {The Coronal Structure of AB Doradus}.
\newblock ApJ, 721(1), 80--89.
\newblock \url{https://doi.org/10.1088/0004-637X/721/1/80}.

\bibitem[{{Dal} and {Tas}(2003)}]{2003IBVS.5390....1D}
{Dal}, H.~A. and {Tas}, G. (2003) {New Photoelectric Photometry of the Young
  Star LO Pegasi}.
\newblock IBVS, 5390, 1.

\bibitem[{{den Herder} et~al.(2001){den Herder}, {Brinkman}, {Kahn},
  {Branduardi-Raymont}, {Thomsen}, {Aarts}, {Audard}, {Bixler}, {den Boggende},
  {Cottam}, {Decker}, {Dubbeldam}, {Erd}, {Goulooze}, {G{\"u}del}, {Guttridge},
  {Hailey}, {Janabi}, {Kaastra}, {de Korte}, {van Leeuwen}, {Mauche},
  {McCalden}, {Mewe}, {Naber}, {Paerels}, {Peterson}, {Rasmussen}, {Rees},
  {Sakelliou}, {Sako}, {Spodek}, {Stern}, {Tamura}, {Tandy}, {de Vries},
  {Welch} and {Zehnder}}]{2001A&A...365L...7D}
{den Herder}, J.~W., {Brinkman}, A.~C., {Kahn}, S.~M., {Branduardi-Raymont},
  G., {Thomsen}, K., {Aarts}, H., {Audard}, M., {Bixler}, J.~V., {den
  Boggende}, A.~J., {Cottam}, J., {Decker}, T., {Dubbeldam}, L., {Erd}, C.,
  {Goulooze}, H., {G{\"u}del}, M., {Guttridge}, P., {Hailey}, C.~J., {Janabi},
  K.~A., {Kaastra}, J.~S., {de Korte}, P.~A.~J., {van Leeuwen}, B.~J.,
  {Mauche}, C., {McCalden}, A.~J., {Mewe}, R., {Naber}, A., {Paerels}, F.~B.,
  {Peterson}, J.~R., {Rasmussen}, A.~P., {Rees}, K., {Sakelliou}, I., {Sako},
  M., {Spodek}, J., {Stern}, M., {Tamura}, T., {Tandy}, J., {de Vries}, C.~P.,
  {Welch}, S. and {Zehnder}, A. (2001) {The Reflection Grating Spectrometer on
  board XMM-Newton}.
\newblock A\&A, 365, L7--L17.
\newblock \url{https://doi.org/10.1051/0004-6361:20000058}.

\bibitem[{{Drake} et~al.(2014){Drake}, {Ratzlaff}, {Kashyap}, {Huenemoerder},
  {Wargelin} and {Pease}}]{2014ApJ...783....2D}
{Drake}, J.~J., {Ratzlaff}, P., {Kashyap}, V., {Huenemoerder}, D.~P.,
  {Wargelin}, B.~J. and {Pease}, D.~O. (2014) {A 33 Yr Constancy of the X-Ray
  Coronae of AR Lac and Eclipse Diagnosis of Scale Height}.
\newblock ApJ, 783(1), 2.
\newblock \url{https://doi.org/10.1088/0004-637X/783/1/2}.

\bibitem[{{Eibe} et~al.(1999){Eibe}, {Byrne}, {Jeffries} and
  {Gunn}}]{1999A&A...341..527E}
{Eibe}, M.~T., {Byrne}, P.~B., {Jeffries}, R.~D. and {Gunn}, A.~G. (1999)
  {Evidence for large-scale, global mass inflow and flaring on the late-type
  fast rotator BD+22 deg 4409}.
\newblock A\&A, 341, 527--538.

\bibitem[{{Haisch} et~al.(1991){Haisch}, {Strong} and
  {Rodono}}]{1991ARA&A..29..275H}
{Haisch}, B., {Strong}, K.~T. and {Rodono}, M. (1991) {Flares on the Sun and
  other stars.}
\newblock ARA\&A, 29, 275--324.
\newblock \url{https://doi.org/10.1146/annurev.aa.29.090191.001423}.

\bibitem[{{Hussain} et~al.(2007){Hussain}, {Jardine}, {Donati}, {Brickhouse},
  {Dunstone}, {Wood}, {Dupree}, {Collier Cameron} and
  {Favata}}]{2007MNRAS.377.1488H}
{Hussain}, G.~A.~J., {Jardine}, M., {Donati}, J.~F., {Brickhouse}, N.~S.,
  {Dunstone}, N.~J., {Wood}, K., {Dupree}, A.~K., {Collier Cameron}, A. and
  {Favata}, F. (2007) {The coronal structure of AB Doradus determined from
  contemporaneous Doppler imaging and X-ray spectroscopy}.
\newblock MNRAS, 377(4), 1488--1502.
\newblock \url{https://doi.org/10.1111/j.1365-2966.2007.11692.x}.

\bibitem[{{Jetsu} et~al.(1994){Jetsu}, {Tuominen}, {Grankin}, {Mel'Nikov} and
  {Schevchenko}}]{1994A&A...282L...9J}
{Jetsu}, L., {Tuominen}, I., {Grankin}, K.~N., {Mel'Nikov}, S.~Y. and
  {Schevchenko}, V.~S. (1994) {New ``flip-flop'' of FK Comae Berenices.}
\newblock A\&A, 282, L9--L12.

\bibitem[{{Johnstone} et~al.(2010){Johnstone}, {Jardine} and
  {Mackay}}]{2010MNRAS.404..101J}
{Johnstone}, C., {Jardine}, M. and {Mackay}, D.~H. (2010) {Modelling stellar
  coronae from surface magnetograms: the role of missing magnetic flux}.
\newblock MNRAS, 404(1), 101--109.
\newblock \url{https://doi.org/10.1111/j.1365-2966.2010.16298.x}.

\bibitem[{{Karmakar} et~al.(2016){Karmakar}, {Pandey}, {Savanov}, {Ta{\c{s}}},
  {Pandey}, {Misra}, {Joshi}, {Dmitrienko}, {Sakamoto}, {Gehrels} and
  {Okajima}}]{2016MNRAS.459.3112K}
{Karmakar}, S., {Pandey}, J.~C., {Savanov}, I.~S., {Ta{\c{s}}}, G., {Pandey},
  S.~B., {Misra}, K., {Joshi}, S., {Dmitrienko}, E.~S., {Sakamoto}, T.,
  {Gehrels}, N. and {Okajima}, T. (2016) {LO Peg: surface differential
  rotation, flares, and spot-topographic evolution}.
\newblock MNRAS, 459(3), 3112--3129.
\newblock \url{https://doi.org/10.1093/mnras/stw855}.

\bibitem[{{Lalitha} et~al.(2017){Lalitha}, {Schmitt} and
  {Singh}}]{2017A&A...602A..26L}
{Lalitha}, S., {Schmitt}, J.~H.~M.~M. and {Singh}, K.~P. (2017) {Structure and
  variability in the corona of the ultrafast rotator LO Pegasi}.
\newblock A\&A, 602, A26.
\newblock \url{https://doi.org/10.1051/0004-6361/201629482}.

\bibitem[{{Lister} et~al.(1999){Lister}, {Collier Cameron} and
  {Bartus}}]{1999MNRAS.307..685L}
{Lister}, T.~A., {Collier Cameron}, A. and {Bartus}, J. (1999) {Doppler imaging
  of BD+22 deg4409 (LO Peg) using least-squares deconvolution}.
\newblock MNRAS, 307(3), 685--694.
\newblock \url{https://doi.org/10.1046/j.1365-8711.1999.02662.x}.

\bibitem[{{Mason} et~al.(2001){Mason}, {Breeveld}, {Much}, {Carter}, {Cordova},
  {Cropper}, {Fordham}, {Huckle}, {Ho}, {Kawakami}, {Kennea}, {Kennedy},
  {Mittaz}, {Pandel}, {Priedhorsky}, {Sasseen}, {Shirey}, {Smith} and
  {Vreux}}]{2001A&A...365L..36M}
{Mason}, K.~O., {Breeveld}, A., {Much}, R., {Carter}, M., {Cordova}, F.~A.,
  {Cropper}, M.~S., {Fordham}, J., {Huckle}, H., {Ho}, C., {Kawakami}, H.,
  {Kennea}, J., {Kennedy}, T., {Mittaz}, J., {Pandel}, D., {Priedhorsky},
  W.~C., {Sasseen}, T., {Shirey}, R., {Smith}, P. and {Vreux}, J.~M. (2001)
  {The XMM-Newton optical/UV monitor telescope}.
\newblock A\&A, 365, L36--L44.
\newblock \url{https://doi.org/10.1051/0004-6361:20000044}.

\bibitem[{{Pandey} et~al.(2009){Pandey}, {Medhi}, {Sagar} and
  {Pandey}}]{2009MNRAS.396.1004P}
{Pandey}, J.~C., {Medhi}, B.~J., {Sagar}, R. and {Pandey}, A.~K. (2009) {LO
  Pegasi: an investigation of multiband optical polarization}.
\newblock MNRAS, 396(2), 1004--1011.
\newblock \url{https://doi.org/10.1111/j.1365-2966.2009.14762.x}.

\bibitem[{{Pandey} and {Singh}(2008)}]{2008MNRAS.387.1627P}
{Pandey}, J.~C. and {Singh}, K.~P. (2008) {A study of X-ray flares - I. Active
  late-type dwarfs}.
\newblock MNRAS, 387(4), 1627--1648.
\newblock \url{https://doi.org/10.1111/j.1365-2966.2008.13342.x}.

\bibitem[{{Pandey} and {Singh}(2012)}]{2012MNRAS.419.1219P}
{Pandey}, J.~C. and {Singh}, K.~P. (2012) {A study of X-ray flares - II. RS
  CVn-type binaries}.
\newblock MNRAS, 419(2), 1219--1237.
\newblock \url{https://doi.org/10.1111/j.1365-2966.2011.19776.x}.

\bibitem[{{Pandey} et~al.(2005){Pandey}, {Singh}, {Drake} and
  {Sagar}}]{2005AJ....130.1231P}
{Pandey}, J.~C., {Singh}, K.~P., {Drake}, S.~A. and {Sagar}, R. (2005) {Optical
  and X-Ray Studies of Chromospherically Active Stars: FR Cancri, HD 95559, and
  LO Pegasi}.
\newblock AJ, 130(3), 1231--1246.
\newblock \url{https://doi.org/10.1086/432539}.

\bibitem[{{Piluso} et~al.(2008){Piluso}, {Lanza}, {Pagano}, {Lanzafame} and
  {Donati}}]{2008MNRAS.387..237P}
{Piluso}, N., {Lanza}, A.~F., {Pagano}, I., {Lanzafame}, A.~C. and {Donati},
  J.~F. (2008) {Doppler imaging of the young late-type star LO Pegasi
  (BD+22{\textdegree}4409) in 2003 September}.
\newblock MNRAS, 387(1), 237--246.
\newblock \url{https://doi.org/10.1111/j.1365-2966.2008.13153.x}.

\bibitem[{{Reale}(2007)}]{2007A&A...471..271R}
{Reale}, F. (2007) {Diagnostics of stellar flares from X-ray observations: from
  the decay to the rise phase}.
\newblock A\&A, 471(1), 271--279.
\newblock \url{https://doi.org/10.1051/0004-6361:20077223}.

\bibitem[{{Savanov} et~al.(2016){Savanov}, {Puzin}, {Dmitrienko}, {Karpov},
  {Beskin}, {Biryukov}, {Bondar}, {Ivano}, {Katkova}, {Orekhova}, {Perkov},
  {Sasyuk}, {Romanyuk}, {Semenko}, {Kudryavtsev}, {Karmakar}, {Pandey},
  {Pandey}, {Joshi} and {Misra}}]{2016AcA....66..381S}
{Savanov}, I.~S., {Puzin}, V.~I., {Dmitrienko}, E.~S., {Karpov}, S.~V.,
  {Beskin}, G.~M., {Biryukov}, A.~V., {Bondar}, S.~F., {Ivano}, E.~A.,
  {Katkova}, E.~V., {Orekhova}, N., {Perkov}, A.~V., {Sasyuk}, V.~V.,
  {Romanyuk}, I.~I., {Semenko}, E.~A., {Kudryavtsev}, D., {Karmakar}, S.,
  {Pandey}, J.~C., {Pandey}, S.~B., {Joshi}, S. and {Misra}, K. (2016)
  {Photometric Observations of LO Peg in 2014-2015}.
\newblock AcA, 66(3), 381--390.

\bibitem[{{Siarkowski}(1992)}]{1992MNRAS.259..453S}
{Siarkowski}, M. (1992) {Three-dimensional deconvolution of X-ray emission from
  AR Lac.}
\newblock MNRAS, 259, 453--456.
\newblock \url{https://doi.org/10.1093/mnras/259.3.453}.

\bibitem[{{Siarkowski} et~al.(1996){Siarkowski}, {Pres}, {Drake}, {White} and
  {Singh}}]{1996ApJ...473..470S}
{Siarkowski}, M., {Pres}, P., {Drake}, S.~A., {White}, N.~E. and {Singh}, K.~P.
  (1996) {Corona(e) of AR Lacertae. II. The Spatial Structure}.
\newblock ApJ, 473, 470.
\newblock \url{https://doi.org/10.1086/178159}.

\bibitem[{{Singh} and {Pandey}(2022)}]{2022ApJ...934...20S}
{Singh}, G. and {Pandey}, J.~C. (2022) {An X-Ray Study of Coronally Connected
  Active Eclipsing Binaries}.
\newblock ApJ, 934(1), 20.
\newblock \url{https://doi.org/10.3847/1538-4357/ac7716}.

\bibitem[{{Str{\"u}der} et~al.(2001){Str{\"u}der}, {Briel}, {Dennerl},
  {Hartmann}, {Kendziorra}, {Meidinger}, {Pfeffermann}, {Reppin}, {Aschenbach},
  {Bornemann}, {Br{\"a}uninger}, {Burkert}, {Elender}, {Freyberg}, {Haberl},
  {Hartner}, {Heuschmann}, {Hippmann}, {Kastelic}, {Kemmer}, {Kettenring},
  {Kink}, {Krause}, {M{\"u}ller}, {Oppitz}, {Pietsch}, {Popp}, {Predehl},
  {Read}, {Stephan}, {St{\"o}tter}, {Tr{\"u}mper}, {Holl}, {Kemmer}, {Soltau},
  {St{\"o}tter}, {Weber}, {Weichert}, {von Zanthier}, {Carathanassis}, {Lutz},
  {Richter}, {Solc}, {B{\"o}ttcher}, {Kuster}, {Staubert}, {Abbey}, {Holland},
  {Turner}, {Balasini}, {Bignami}, {La Palombara}, {Villa}, {Buttler},
  {Gianini}, {Lain{\'e}}, {Lumb} and {Dhez}}]{2001A&A...365L..18S}
{Str{\"u}der}, L., {Briel}, U., {Dennerl}, K., {Hartmann}, R., {Kendziorra},
  E., {Meidinger}, N., {Pfeffermann}, E., {Reppin}, C., {Aschenbach}, B.,
  {Bornemann}, W., {Br{\"a}uninger}, H., {Burkert}, W., {Elender}, M.,
  {Freyberg}, M., {Haberl}, F., {Hartner}, G., {Heuschmann}, F., {Hippmann},
  H., {Kastelic}, E., {Kemmer}, S., {Kettenring}, G., {Kink}, W., {Krause}, N.,
  {M{\"u}ller}, S., {Oppitz}, A., {Pietsch}, W., {Popp}, M., {Predehl}, P.,
  {Read}, A., {Stephan}, K.~H., {St{\"o}tter}, D., {Tr{\"u}mper}, J., {Holl},
  P., {Kemmer}, J., {Soltau}, H., {St{\"o}tter}, R., {Weber}, U., {Weichert},
  U., {von Zanthier}, C., {Carathanassis}, D., {Lutz}, G., {Richter}, R.~H.,
  {Solc}, P., {B{\"o}ttcher}, H., {Kuster}, M., {Staubert}, R., {Abbey}, A.,
  {Holland}, A., {Turner}, M., {Balasini}, M., {Bignami}, G.~F., {La
  Palombara}, N., {Villa}, G., {Buttler}, W., {Gianini}, F., {Lain{\'e}}, R.,
  {Lumb}, D. and {Dhez}, P. (2001) {The European Photon Imaging Camera on
  XMM-Newton: The pn-CCD camera}.
\newblock A\&A, 365, L18--L26.
\newblock \url{https://doi.org/10.1051/0004-6361:20000066}.

\bibitem[{{Turner} et~al.(2001){Turner}, {Abbey}, {Arnaud}, {Balasini},
  {Barbera}, {Belsole}, {Bennie}, {Bernard}, {Bignami}, {Boer}, {Briel},
  {Butler}, {Cara}, {Chabaud}, {Cole}, {Collura}, {Conte}, {Cros}, {Denby},
  {Dhez}, {Di Coco}, {Dowson}, {Ferrando}, {Ghizzardi}, {Gianotti}, {Goodall},
  {Gretton}, {Griffiths}, {Hainaut}, {Hochedez}, {Holland}, {Jourdain},
  {Kendziorra}, {Lagostina}, {Laine}, {La Palombara}, {Lortholary}, {Lumb},
  {Marty}, {Molendi}, {Pigot}, {Poindron}, {Pounds}, {Reeves}, {Reppin},
  {Rothenflug}, {Salvetat}, {Sauvageot}, {Schmitt}, {Sembay}, {Short},
  {Spragg}, {Stephen}, {Str{\"u}der}, {Tiengo}, {Trifoglio}, {Tr{\"u}mper},
  {Vercellone}, {Vigroux}, {Villa}, {Ward}, {Whitehead} and
  {Zonca}}]{2001A&A...365L..27T}
{Turner}, M.~J.~L., {Abbey}, A., {Arnaud}, M., {Balasini}, M., {Barbera}, M.,
  {Belsole}, E., {Bennie}, P.~J., {Bernard}, J.~P., {Bignami}, G.~F., {Boer},
  M., {Briel}, U., {Butler}, I., {Cara}, C., {Chabaud}, C., {Cole}, R.,
  {Collura}, A., {Conte}, M., {Cros}, A., {Denby}, M., {Dhez}, P., {Di Coco},
  G., {Dowson}, J., {Ferrando}, P., {Ghizzardi}, S., {Gianotti}, F., {Goodall},
  C.~V., {Gretton}, L., {Griffiths}, R.~G., {Hainaut}, O., {Hochedez}, J.~F.,
  {Holland}, A.~D., {Jourdain}, E., {Kendziorra}, E., {Lagostina}, A., {Laine},
  R., {La Palombara}, N., {Lortholary}, M., {Lumb}, D., {Marty}, P., {Molendi},
  S., {Pigot}, C., {Poindron}, E., {Pounds}, K.~A., {Reeves}, J.~N., {Reppin},
  C., {Rothenflug}, R., {Salvetat}, P., {Sauvageot}, J.~L., {Schmitt}, D.,
  {Sembay}, S., {Short}, A.~D.~T., {Spragg}, J., {Stephen}, J., {Str{\"u}der},
  L., {Tiengo}, A., {Trifoglio}, M., {Tr{\"u}mper}, J., {Vercellone}, S.,
  {Vigroux}, L., {Villa}, G., {Ward}, M.~J., {Whitehead}, S. and {Zonca}, E.
  (2001) {The European Photon Imaging Camera on XMM-Newton: The MOS cameras}.
\newblock A\&A, 365, L27--L35.
\newblock \url{https://doi.org/10.1051/0004-6361:20000087}.

\end{thebibliography}

\end{document}